\documentclass{elsarticle}
\usepackage{amsmath,amssymb}
\usepackage{lscape}

\begin{document}
\title{Search for Photon-Photon Elastic Scattering\\in the X-ray Region}
\author[phys]{T.~Inada}
\ead{tinada@icepp.s.u-tokyo.ac.jp}
\author[phys]{T.~Yamaji}
\ead{yamaji@icepp.s.u-tokyo.ac.jp}
\author[phys]{S.~Adachi}
\author[icepp]{T.~Namba}
\author[phys]{S.~Asai}
\author[icepp]{T.~Kobayashi}
\author[riken]{\\K.~Tamasaku}
\author[riken]{Y.~Tanaka}
\author[riken]{Y.~Inubushi}
\author[riken]{K.~Sawada}
\author[riken]{M.~Yabashi}
\author[riken]{T.~Ishikawa}
\address[phys]{Department of Physics, Graduate School of Science, The University of Tokyo, 7-3-1 Hongo, Bunkyo, Tokyo 113-0033, Japan}
\address[icepp]{International Center for Elementary Particle Physics, The University of Tokyo, 7-3-1 Hongo, Bunkyo, Tokyo 113-0033, Japan}
\address[riken]{RIKEN SPring-8 Center, 1-1-1 Kouto, Sayo-cho, Sayo-gun, Hyogo 679-5148, Japan}

\begin{abstract}
We report the first results of a search for real photon-photon scattering using X rays.
A novel system is developed to split and collide X-ray pulses by applying interferometric techniques. 
A total of $6.5\times10^{5}$ pulses (each containing about $10^{11}$ photons) from an X-ray Free-Electron Laser
are injected into the system.
No scattered events are observed, and
an upper limit of $1.7\times 10^{-24}$ ${\rm m^{2}}$ (95\% C.L.) is obtained on the photon-photon elastic scattering cross section at 6.5 keV.
\end{abstract}

\maketitle

\section{Introduction}
Photon-photon ($\gamma$-$\gamma$) scattering is a pure quantum-electrodynamical (QED) process, in which the box diagram with an electron gives the leading-order contribution.
This QED process was predicted in 1933\cite{cite_halpern} as a characteristic 
effect of the non-linearity of the quantum vacuum.
Although indirect tests have been performed using Delbr\"{u}ck scattering\cite{cite_delbruck},
$\gamma$-$\gamma$ scattering of real photons has not yet been directly observed.
The direct observation of this process would provide solid evidence for vacuum polarization caused by virtual electrons,
and is considered 
an ultimate test of QED.

In the past, visible lasers have been used to search for real $\gamma$-$\gamma$ scattering\cite{cite_visible1, cite_visible2}.
At low energies ($\hbar\omega \ll m_{{\rm e}}c^{2}$), however, the QED cross section of 
$\gamma$-$\gamma$ scattering is suppressed by the sixth power of the ratio $\hbar \omega_/m_{{\rm e}}c^{2}$,
where $\omega$ is the photon energy and $m_{{\rm e}}$ the electron mass\cite{cite_cross_section}.
A significant enhancement of the cross section can be obtained by using higher-energy photons, for example, X rays.

In addition to the context of QED tests, studies in a different energy region are important for the search for new physics beyond the Standard Model.
A theoretical benchmark for such processes is s-channel production of light pseudoscalar bosons, namely axions\cite{cite_axion}.
If axions exist with a mass in the X-ray region, invisible axion models\cite{cite_axion_model} predict that $\gamma$-$\gamma$ scattering is dominated by the contribution from axions under a resonant condition\cite{cite_axion_gamgam}.

In this paper we report the first results of a direct search for $\gamma$-$\gamma$ scattering in the X-ray region.
Our experiment is performed at the SPring-8 Angstrom Compact Free-Electron Laser\cite{cite_sacla} (SACLA),
an X-ray Free-Electron Laser (XFEL) which provides one of the 
world's
most brilliant X-ray sources.
To ensure the collision of $\gamma$-$\gamma$ in both space and time,
a high-precision collision system is developed, using a single silicon crystal to split and collide X-ray pulses.

The differential cross section of the scattering of two photons 
with the same linear polarization is\cite{cite_cross_section}
\begin{equation}
\left(\frac{{\rm d}\sigma_{\gamma\gamma\rightarrow\gamma\gamma}}{{\rm d}\Omega}\right)_{\rm {QED}}
=\frac{\alpha^{4}\omega_{{\rm CMS}}^{6}}{\left(180\pi\right)^{2}m_{{\rm e}}^{8}}
\left(260\cos^{4}{\rm \phi}+328\cos^{2}{\rm \phi}+580\right),
\label{eq0}
\end{equation}
where $\alpha$ is the fine structure constant and $\phi$ is the scattering angle in the center of mass system (CMS).
The CMS photon energy ($\omega_{{\rm CMS}}$) is 6.5 keV in our system, giving a cross section  
around 23 orders of magnitude larger than that at 1 eV, the typical energy of visible lasers.

\section{Experimental Setup}

\begin{figure}
\begin{center}
\includegraphics[width=0.9\textwidth]{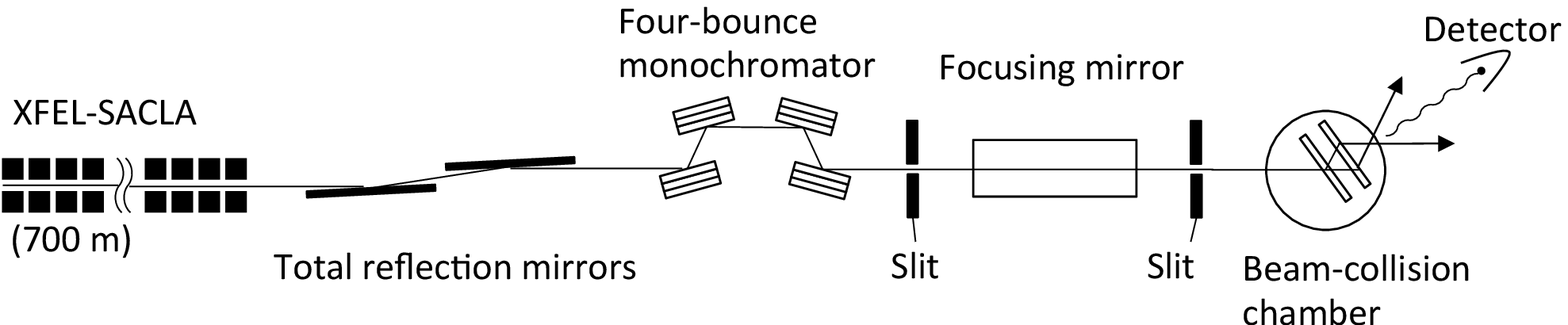}
\end{center}
\caption{Side view of the beamline components.}
\label{fig_schematic}
\end{figure}

Figure~\ref{fig_schematic} shows a schematic view of the experimental setup.
A pulsed beam of synchrotron radiation (repetition rate of 20 Hz) is produced in the BL(beamline)-3 of the SACLA facility (total length of 700 m),
which is mainly composed of a low-emittance electron gun, C-band high gradient accelerators, and nineteen undulators.
Electron beam energy and vertical gaps in the undulator magnets are tuned to produce X rays with an energy of $\omega_{0}=$10.985 keV.
The pulse length of the beam is measured to be less than 10 fs\cite{cite_time}, and each pulse contains about $10^{11}$ photons.
The beam from the undulator is linearly polarized in the horizontal direction.

Higher harmonics of the X-ray beam are contained in the beam after the undulators.
These higher components are removed by a pair of carbon-coated mirrors\cite{cite_mirror} placed in an optics hutch downstream of an undulator building.
Their glancing angle is set to 2.0 mrad to give total reflection of X rays with an energy below 11 keV.
The angle of both mirrors are the same, giving parallel incoming and outgoing beams.

The energy spectrum of the beam after the undulators has a bandwidth of about 50 eV.
The monochromaticity is improved to the level of $\sim$ 60 meV (FWHM) by a four-bounce monochromator with silicon (440) reflection,
placed downstream of the total reflection mirrors.
The beams are reflected and filtered in energy with the Bragg condition at the first surface.
The second one is used to change the beam axis.
The same mechanism is used in a second pair of silicon surfaces, restoring the vertical offset of the beam axis.

The horizontal beam size is focused from $\sim$ 200 $\mu$m to $\sim$ 1 $\mu$m by an elliptical focusing mirror (focusing length=1.5 m, Rayleigh length$\sim$10 mm)\cite{cite_kb}.
The vertical direction is not focused and remains at its original size ($\sim200$ $\mu$m),
since the vertical component of the beam contributes to the Bragg reflection at the silicon splitter.
The spatial beam profile has a Gaussian distribution.
In order to improve the beam quality, a pair of four-jar slits is inserted along the beam axis, on either side of the focusing mirror.
These slits are used to block the side-lobes of the beam and diffuse scattering from the mirror.

\begin{figure}
\begin{center}
\includegraphics[width=0.9\textwidth]{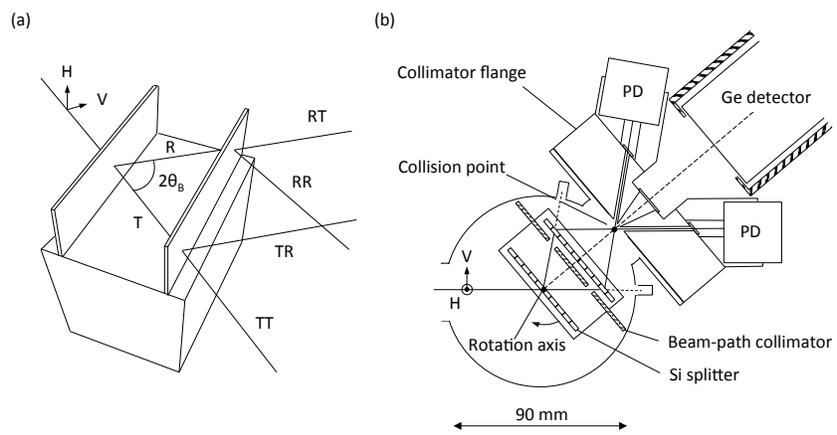}
\end{center}
\caption{(a)Four beams after the two blades of the silicon beam splitter.
Horizontal (H) and vertical (V) directions are shown in a plane which is normal to the beam axis.
(b)Magnified view of the beam-collision chamber.}
\label{fig_collision}
\end{figure}

An X-ray interferometer technique\cite{cite_laue} is applied to split and collide X-ray pulses.
A schematic view of the beam splitter is shown in Fig.~\ref{fig_collision}a.
The collision of the beams is assured by the geometrical equivalence of the beam paths split by symmetric Laue reflection\cite{cite_cross}.
The beams are split and collided by two 0.6~mm-thick blades of (440) silicon crystal cut from a high-quality single crystal. 
The blades are separated by 25 mm.
A magnified view of the beam-collision chamber is shown in Fig.~\ref{fig_collision}b.
The two blades are placed so as to split and collide X rays in a vertical-longitudinal plane; this geometry minimizes the loss 
of the $\pi$-polarized component.
The blades are installed in a cylindrical vacuum chamber (beam-collision chamber) and
fixed to a rotary feedthrough connected to a goniometer.
The center of the first blade is aligned to coincide with the rotation axis of the goniometer.
The rotation angle of the lattice plane (${\rm \theta_{B}}$) is tuned to satisfy the condition for Laue reflection.

The incident beams are split into two coherent beams at the first blade; one is the transmitted beam (T) and the other is the reflected one (R), with a reflection angle of 2${\rm \theta_{B}}=72^\circ$ (Fig.~\ref{fig_collision}a).
A similar process occurs at the second blade, creating four beams after the two blades.
Two beams (TT and RT) are dumped into pits on the wall of the chamber.
The other beams (TR and RR) collide at a collision point and pass through a vacuum window made of 125 ${\rm \mu}$m-thick polyimide.
The collision point is adjusted to be at the focusing point of the mirror.
The intensities of the two beams are monitored pulse-by-pulse by silicon PIN photodiodes (PDs, HAMAMATSU S3590-09),
inserted in stainless steel holders fixed to the chamber flange and wrapped in a 3 mm-thick lead sheet to avoid photon leakage.
In order to reduce X rays scattered by air, the beam path is evacuated upstream of the collision chamber ($<10^{-5}$ Pa) and inside the chamber ($<10^{-2}$ Pa).

The TR and RR beams collide obliquely with a crossing angle of $2{\rm \theta_{B}}$.
The CMS energy of each photon is $\omega_{0}\sin{\rm \theta_{B}}$
and the two-photon system is boosted with an energy of $2\omega_{0}\cos{\rm \theta_{B}}$.
The maximum and minimum energies of scattered photons, given by $\omega_{0}\left(1\pm\cos{\rm \theta_{B}}\right)$, 
occur for photons scattered along the boost axis.
Since forward-scattered photons have higher energy than the original photons,
it is easy to separate their signal from backgrounds.
To select forward-scattered photons, a collimator flange with a 
conically tapered hole is placed after the collision point.
The full angle of the cone is 25$^{\circ}$, defining the signal energy region from 18.1 keV to 19.9 keV.
A second beam-path collimator (stainless steel, thickness=2 mm) with two holes (diameter=6 mm) is inserted between the two blades.
The two beams from the first blade pass through the collimator, while
Rayleigh- or Compton-scattered photons from the blade are blocked.

A large-volume germanium detector (CANBERRA BE2825) is used to detect the signal X rays.
The diameter and thickness of the crystal are 60 mm and 25 mm, respectively.
The detector is tilted by 36$^\circ$ with respect to the beam axis, and the center of the crystal is aligned to the axis of the tapered collimator hole, as shown in Fig.~\ref{fig_collision}b.
The detector is shielded by 5 mm-thick brass on its curved surface, and 2 mm-thick stainless steel, with a circular opening
of diameter 80 mm, much larger than the projection of the taper cone, on the front surface.
One output from the detector's preamplifier is shaped by an amplifier, whose output is measured by a peak hold ADC.
A second preamplifier output is fed to a 8-bit flash ADC, with sampling rate of 50 MHz and gate width of 20 $\mu$s, to record the waveform.
Both ADCs are gated by triggers synchronized to the X-ray pulses.
An energy calibration is performed at the primary beam energy $\omega_{0}$, and using a $^{57}$Co source during the measurement.
The energy resolution and detection efficiency ($\epsilon_{d}$) are measured using 
${\rm ^{55}Fe}$, ${\rm ^{57}Co}$, ${\rm ^{68}Ge}$ and ${\rm ^{241}Am}$ sources;
typical values at the lower edge of the signal region are 0.18 keV (standard deviation) and 67\%,
respectively\cite{cite_paraphoton}.

\section{Measurement and Analysis}

The experiment has been performed on 23rd and 24th July 2013 at SACLA.
The data acquisition time is 9 hours, corresponding to a total injection of $6.5\times10^{5}$ pulses.
A timing cut is applied to select events within $\pm 1$ ${\rm \mu}$s of the arrival of X-ray pulses.
The unsynchronized count rate is $(1.1\pm0.1)\times10^{-4}$ pulse$^{-1}$, in agreement with environmental background counts.

\begin{figure}
\hspace{-25mm}
\begin{center}
\includegraphics[width=0.9\textwidth]{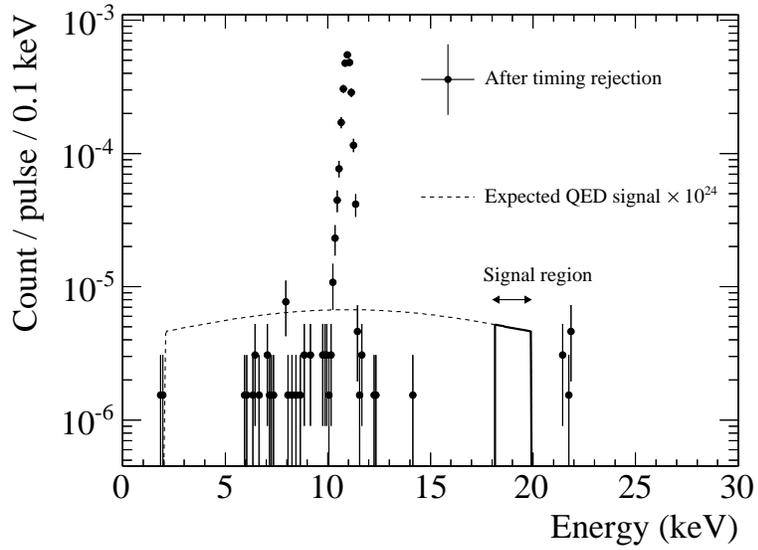}
\end{center}
\caption{
Energy spectrum obtained from the total injection of $6.5\times10^{5}$ pulses, after the timing rejection cut.
The dotted line shows the expected QED signal shape in the full solid angle (signal strength is $\times10^{24}$).
The solid line corresponds to the signal region from 18.1 keV ($\theta=12.5^{\circ}$) to 19.9 keV ($\theta=0$).
}
\label{fig_spectrum}
\end{figure}

Figure~\ref{fig_spectrum} shows the energy spectrum after the timing cut.
A peak from Rayleigh-scattered stray X rays is observed around the beam energy $\omega_{0}$,
and some events are also observed at 2$\omega_{0}$ due to the pileup of two photons.
These count rates are $(2.6\pm0.1)\times10^{-3}$ pulse$^{-1}$ and $(9.2\pm3.8)\times10^{-6}$ pulse$^{-1}$, respectively.
The measured pileup rate is consistent with the expected rate of $(6.8\pm0.3)\times10^{-6}$ pulse$^{-1}$.
Additional events are observed below the beam energy,
corresponding to X-ray fluorescence from elements in the stainless-steel (mainly Cr, Fe, Co and Ni) of the chamber and collimators.
Their count rate is measured to be $(3.4\pm0.7)\times10^{-5}$ pulse$^{-1}$.

No events are observed in the signal region (18.1 keV - 19.9 keV).
The expected background rate in the signal region is calculated as the pileup of fluorescence and primary photons,
giving an expected count rate of $(8.8\pm1.9)\times10^{-8}$ pulse$^{-1}$, consistent with the null observation.
The upper limit on the signal ($N_{{\rm UL}}$) is calculated to be $4.6\times10^{-6}$ pulse$^{-1}$ at 95\% C.L.

The luminosity of the colliding X-ray pulses is given by\cite{luminosity}
\begin{equation}
L=
\frac{N_{1}N_{2}}{4\pi\ell_{{\rm v}}\ell_{{\rm h}}
\sqrt{ 1+
\left(\frac{\ell_{\rm t}}{\ell_{{\rm v}}}\right)^{2}
\tan^{2}\frac{{\rm \theta_{c}}}{2}
}}
\label{eq1}
\end{equation}
where $N_{1}$ and $N_{2}$ are numbers of photons per pulse in the TR and RR beams,
$\theta_{{\rm c}}$ is the crossing angle,
and $\ell_{{\rm v}}$, $\ell_{{\rm h}}$ and $\ell_{{\rm t}}$ are the beam widths in the vertical, horizontal and time directions, respectively.
$N_{1}$ and $N_{2}$ are monitored by the PDs, and have mean values of $2.0\times10^{5}$ pulse$^{-1}$ and $1.7\times10^{5}$ pulse$^{-1}$, respectively.
Their mean product, calculated from the pulse-by-pulse product of $N_{1}N_{2}$, is $5.5\times10^{10}$ pulse$^{-1}$.

The beam profile is measured using an edge scan method.
$\ell_{{\rm h}}$ is measured by a vertical gold wire (diameter=200 $\mu$m).
The wire is inserted at the focusing point and moved along the horizontal direction.
The observed beam profile is consistent with a Gaussian distribution of width $0.78\pm0.08$(sys.) $\mu$m.
The measurement is repeated three times, and their standard deviation is taken into account as a systematic error,
since the focused width is comparable to an accuracy of the measurement.
The vertical width ($\ell_{{\rm v}}$) is measured using the tantalum blade of a four-jar slit as an edge.
The widths of the TR and RR beams are not identical since they undergo different interactions at the silicon blades.
The measured widths are $215\pm8$ $\mu$m for the TR beam ($\ell_{{\rm vTR}}$) and $200\pm4$ $\mu$m for the RR beam ($\ell_{{\rm vRR}}$).
The average value of $\ell_{{\rm v}}$ is calculated as $\sqrt{(\ell_{{\rm vTR}}^{2}+\ell_{{\rm vRR}}^{2})/2}$.
The pulse length $\ell_{{\rm t}}$ is so small ($<$ 3 $\mu$m\cite{cite_time}) compared to $\ell_{{\rm v}}$ that the term containing the crossing angle in Eq.~\ref{eq1} becomes negligible.
The luminosity is calculated to be $(2.1\pm0.3)\times10^{19}$ m$^{-2}$$\cdot$pulse$^{-1}$.
A value one standard deviation smaller than the central value is conservatively used in the analysis.

The limit on the QED cross section of $\gamma$-$\gamma$ elastic scattering $\sigma_{\gamma\gamma\rightarrow\gamma\gamma}$ is calculated by
\begin{equation}
\epsilon_{d} \epsilon_{c} \sigma_{\gamma\gamma\rightarrow\gamma\gamma} L = N_{{\rm UL}},
\end{equation}
where $\epsilon_{c}$ is the signal coverage, 
determined from Eq.~\ref{eq0} and the detector acceptance ($0<\theta<12.5^{\circ}$) to be 0.18.
The minimum value of $\epsilon_{d}=67$\% is conservatively used for the limit calculation.
The limit is calculated to be
\begin{equation}
\sigma_{\gamma\gamma\rightarrow\gamma\gamma}<1.7\times10^{-24}\ {\rm m^{2}}
\end{equation}
at 95\% C.L.

\begin{figure}
\hspace{-25mm}
\begin{center}
\includegraphics[width=0.9\textwidth]{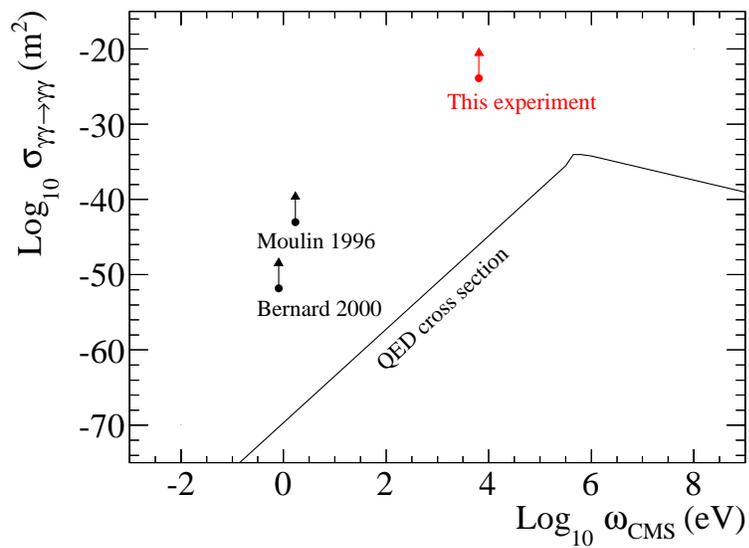}
\end{center}
\caption{
95\% C.L. upper limit on the QED cross section of elastic $\gamma$-$\gamma$ scattering obtained with XFEL-SACLA.
QED prediction and limits from other experiments are also shown.
Moulin\cite{cite_visible1} is a visible laser experiment using a head-on collision geometry.
Bernard\cite{cite_visible2} is an experiment applying four-wave induced mixing with three laser beams.
}
\label{fig_limit}
\end{figure}

Figure~\ref{fig_limit} shows limits on $\sigma_{\gamma\gamma\rightarrow\gamma\gamma}$ obtained by several experiments at different photon energies.
The two points at $\omega_{{\rm CMS}}\sim {\rm eV}$ are results using visible lasers\cite{cite_visible1, cite_visible2}, and
the solid line shows the QED prediction.
Our result is shown at $\omega_{{\rm CMS}}=6.5$ keV.
This is the first constraint on $\sigma_{\gamma\gamma\rightarrow\gamma\gamma}$ in the X-ray region.

\section{Conclusion and prospect}

The first search for elastic $\gamma$-$\gamma$ scattering in the X-ray region has been performed using an XFEL.
A germanium detector is used to search for photons scattered elastically along the boost axis of the two photon system.
No signal events are observed.
The obtained limit on the cross section of $\gamma$-$\gamma$ elastic scattering at a photon energy of 6.5 keV is 
$\sigma_{\gamma\gamma\rightarrow\gamma\gamma}<1.7\times10^{-24}$ m$^{2}$ at 95\% C.L.

Based on the the present experiment, major upgrades of both the XFEL and the detector are underway.
The present mechanism of XFEL amplification is Self-Amplified Spontaneous Emission\cite{cite_sase} (SASE),
the stochastic nature of which limits the bandwidth to $\sim$ 50 eV.
A method to improve the monochromaticity, 
known as the seeding technique\cite{cite_seed}, will be introduced to SACLA in 2014.
Reducing the monochromatic loss yields X-ray pulses with higher intensity, 
and a significant (around $10^{6}$) improvement on the limit is expected.
The high-intensity pulses will give an increase of background events.
In order to reduce background counts,
a pixelized semiconductor detector is currently being studied to replace the present monolithic germanium detector.
The information of hit positions will help discriminate signal from pileup events.

\section*{Acknowledgements}

The experiment was performed at SACLA with the approval of JASRI and the Program Review Committee (Proposal No.~2013A8004).
Commissioning run at BL19LXU in SPring-8 is also performed with the approval of RIKEN (Proposal No.~20120088).
Sincere gratitude is expressed to SACLA engineering team.
We also thank D. Jeans for useful discussions.
Works of T. Inada, T. Yamaji and S. Adachi are supported in part by Advanced Leading Graduate Course for Photon Science (ALPS) at the University of Tokyo.
This research is funded in part by the Japan Society for the Promotion of Science (Grant number 13J07172).

\clearpage

\clearpage

\clearpage

\clearpage

\end{document}